\documentclass[12pt]{article}
\pdfoutput=1
\usepackage{jcappub}

\usepackage{amsmath,amsfonts,amssymb,mathrsfs,graphicx,color,longtable}
\usepackage{hyperref}
\usepackage{graphicx}
\usepackage{amsfonts,amsmath,amssymb,bm}
\usepackage{array}
\usepackage{color}
\usepackage{enumitem}
\usepackage[explicit]{titlesec}
\usepackage[utf8]{inputenc}
\usepackage[normalem]{ulem}
\usepackage{xcolor}

\hypersetup{
     colorlinks   = true,
     citecolor    = red,
     urlcolor     = blue,
     linkcolor    = blue
}

\begin{document}
\title{On the primordial origin of the smoothing excess in the $Planck$ temperature power spectrum in light of LSS data}

\author[1,2,3,4,5,6]{Mario Ballardini,}
\author[4,5]{Fabio Finelli}

\affiliation[1]{Dipartimento di Fisica e Scienze della Terra, Universit\`a degli Studi di Ferrara, via Giuseppe Saragat 1, I-44122 Ferrara, Italy}
\affiliation[2]{INFN, Sezione di Ferrara, via Giuseppe Saragat 1, I-44122 Ferrara, Italy}
\affiliation[3]{Dipartimento di Fisica e Astronomia, Alma Mater Studiorum Universit\`a di Bologna, via Gobetti 93/2, I-40129 Bologna, Italy}
\affiliation[4]{INAF/OAS Bologna, via Piero Gobetti 101, I-40129 Bologna, Italy}
\affiliation[5]{INFN, Sezione di Bologna, via Irnerio 46, I-40126 Bologna, Italy}
\affiliation[6]{Department of Physics \& Astronomy, University of the Western Cape, Cape Town 7535, South Africa}

\emailAdd{mario.ballardini@unife.it}
\emailAdd{fabio.finelli@inaf.it}

\abstract{
The {\em Planck} DR3 measurements of the temperature and polarization anisotropies power spectra of the cosmic microwave background (CMB) show 
an excess of smoothing of the acoustic peaks with respect to $\Lambda$CDM, often quantified by a 
phenomenological parameter $A_{\rm L}$. A specific feature superimposed to the primordial power 
spectrum has been suggested as a physical solution for this smoothing excess.
Here, we investigate the impact of this specific localized oscillation with a frequency linear 
in the wavenumber, designed to mimic the smoothing of CMB temperature spectrum corresponding to 
$A_{\rm L} \simeq 1.1-1.2$ on the matter power spectrum.
We verify the goodness of the predictions in perturbation theory at next-to-leading order with a set of N-body simulations, a necessary step to study the non-linear damping of these primordial oscillations.
We show that for a large portion of the parameter space, the amplitude of this primordial 
oscillation can be strongly damped on the observed nonlinear matter power spectrum at $z=0$, but 
a larger signal is still persistent at $z \lesssim 2$ and is therefore a target for future galaxy surveys at high redshifts. From an analysis of the BOSS DR12 two-point correlation function, we find 
${\cal A}_{\rm lin} < 0.26$ at 95\% CL by keeping the frequency fixed to the best-fit of {\em Planck} data.}

\maketitle

\section{Introduction}
An excess of smoothing in the region of the acoustic peaks of the CMB anisotropy temperature 
power spectrum has been found in all three {\em Planck} data releases 
\cite{Planck:2013pxb,Planck:2015fie,Planck:2018vyg}. This effect is often quantified by a 
phenomenological parameter $A_{\rm L}$ which scales the theoretical prediction for the lensing 
contribution. The excess from $A_{\rm L} = 1$ (the prediction of general relativity) represents 
one of the current intriguing discrepancies within {\em Planck} data plaguing the standard 
$\Lambda$CDM model \cite{Handley:2019tkm,DiValentino:2019qzk,Efstathiou:2020wem}.

Results from the {\em Planck} Collaboration based on DR3 correspond to 
$A_{\rm L} = 1.180 \pm 0.065$ ($2.8\sigma$) using the baseline {\tt Plik} likelihood 
\cite{Planck:2018vyg,Planck:2019nip} and using the {\tt CamSpec} likelihood 
$A_{\rm L} = 1.149 \pm 0.072$ ($2.1\sigma$) \cite{Planck:2018vyg,Efstathiou:2019mdh} both at 68\% CL 
combining low-$\ell$ and high-$\ell$ temperature and polarization data when CMB lensing information 
is not included (see Ref.~\cite{Calabrese:2008rt} for an early indication for higher lensing 
contribution from the combination of WMAP 3Y and ACBAR CMB data). 
Latest results based on the {\tt CamSpec} likelihood with {\em Planck} DR4 also show an excess, $A_{\rm L} = 1.095 \pm 0.056$, although at a lower significance ($1.7\sigma$) mainly due to 
differences in EE and TE \cite{Rosenberg:2022sdy}.
Current results from CMB ground-base telescope are more  consistent with $\Lambda$CDM predictions: 
$A_{\rm L} = 1.01 \pm 0.11$ with ACT DR4 \cite{ACT:2020gnv} and $A_{\rm L} = 0.98 \pm 0.12$ 
with SPT-3G 2018 EE/TE data \cite{SPT-3G:2021eoc}. 
By folding in {\em Planck} CMB lensing data the estimate of $A_{\rm L}$ decreases to smaller values \cite{Planck:2018lbu}.

The residuals between {\em Planck} data and the $\Lambda$CDM best-fit yield an oscillatory 
pattern, which can be mimicked by a specific localized oscillations superimposed to the 
primordial power spectrum (PPS), see 
Refs.~\cite{Domenech:2019cyh,Planck:2018jri,Domenech:2020qay,Hazra:2022rdl,Antony:2022ert}.\footnote{Note that the degeneracy between the effect of lensing on the CMB temperature power spectrum 
and oscillating primordial feature was already pointed out in Ref.~\cite{Hazra:2014jwa}.} 
It has been recently shown that with this type of primordial feature, a higher value for $H_0$ 
and a smaller value of $S_8$ than in $\Lambda$CDM can be simultaneously found 
\cite{Hazra:2022rdl,Antony:2022ert}.

However, oscillations in the PPS cannot mimic $A_{\rm L}$ in all the different CMB fields at 
the same time because the peaks in CMB polarization are out of phase with those in temperature 
\cite{Planck:2018jri}.
Indeed, the phase of the E-mode polarization peaks follows the velocity fluid making the turning 
points of temperature peaks corresponding to zero points of velocity (this implies a $\pi/2$ 
shift in phase).
In addition, the correlation power spectrum TE, being the product of the two, exhibits acoustic 
peaks with twice the acoustic frequency. Moreover, matter and radiation oscillations are not in 
phase implying a $\pi/2$ shift in phase between Fourier mode connected to the decoupling of 
photons and baryon \cite{Eisenstein:1997ik}.
These differences point to the possibility to use future CMB polarization and/or large-scale 
structure (LSS) clustering measurements in order to further test the primordial origin of the 
CMB smoothing excess. 

LSS clustering information has been highlighted as an excellent observable to test primordial 
oscillations on small scales and for high frequencies as shown in many forecast studies in 
Refs.~\cite{Huang:2012mr,Chen:2016zuu,Chen:2016vvw,Ballardini:2016hpi,Ballardini:2017qwq,Slosar:2019gvt,Beutler:2019ojk,Ballardini:2019tuc} and demonstrated on real data in 
Refs.~\cite{Zeng:2018ufm,Beutler:2019ojk,Ballardini:2022wzu}.
Indeed, the case of primordial features with linear and logarithmic oscillations have been 
workhorses for the studies with perturbation theory 
\cite{Vlah:2015zda,Beutler:2019ojk,Vasudevan:2019ewf,Ballardini:2019tuc,Chen:2020ckc,Li:2021jvz}
and N-body simulations \cite{Vlah:2015zda,Ballardini:2019tuc,Chen:2020ckc,Li:2021jvz}
in order to have accurate theoretical predictions for LSS clustering observables.

\begin{figure*}
\centering
\includegraphics[width=\textwidth]{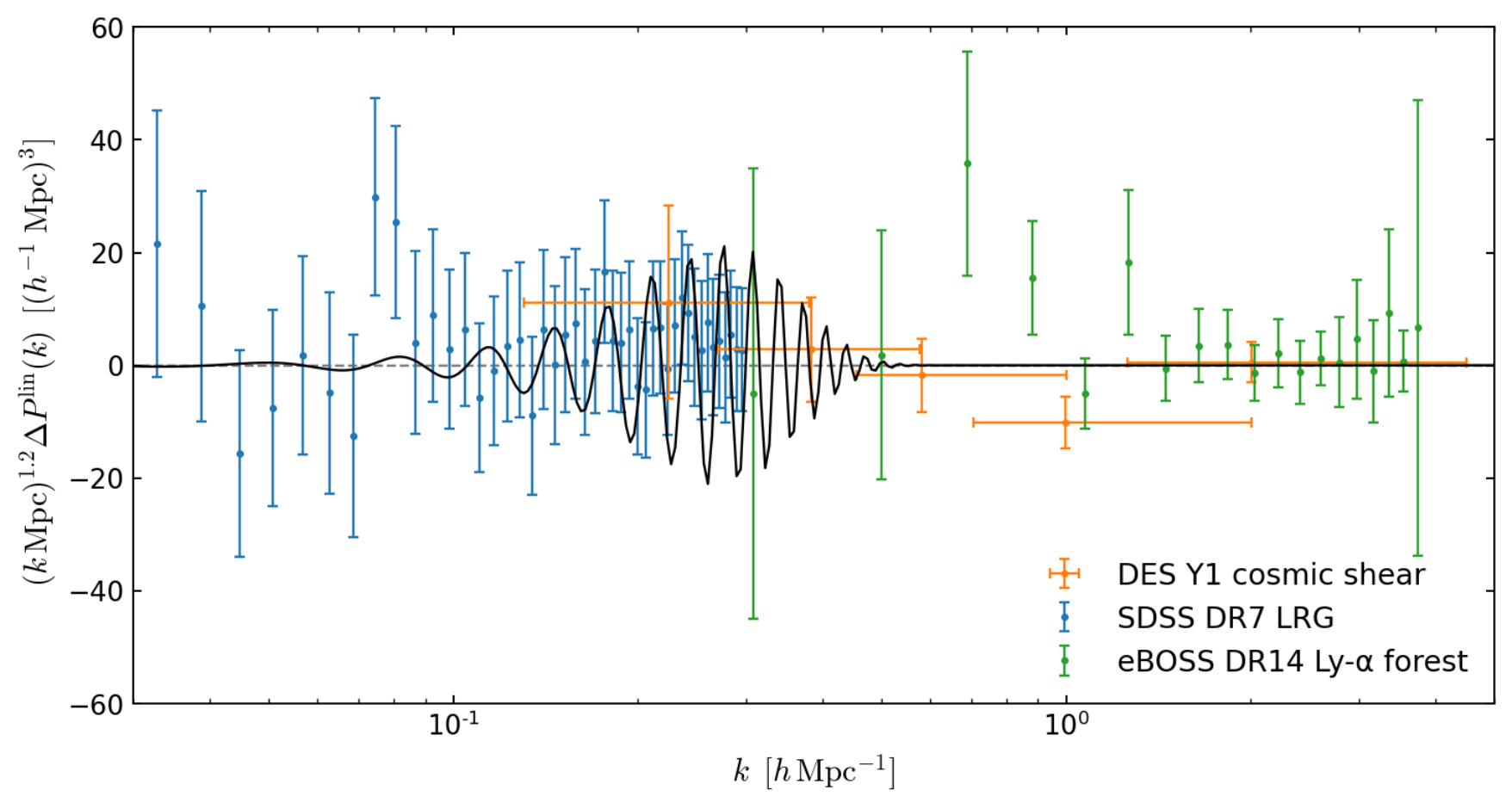}
\caption{Figure comparing measurements \cite{DES:2017qwj,Reid:2009xm,SDSS:2017bih} 
and linear theoretical predictions adapted from Ref.~\cite{Chabanier:2019eai}.}
\label{fig:Pk_error}
\end{figure*}

Indeed, Fig.~\ref{fig:Pk_error} shows that current measurements of the matter power spectrum 
(for a compilation of inferences of the linear matter power spectrum at redshift $z = 0$ 
following the method proposed in Refs.~\cite{Chabanier:2019eai,Tegmark:2002cy}) are potentially 
able to put some constraint on this specific template considering the best-fit amplitude and 
the location of the feature of one of the primordial feature model proposed in order to recover 
$A_{\rm L} \simeq 1$ in Ref.~\cite{Planck:2018jri}.

In this paper, we extensively study the imprint on the matter power spectrum of a localized 
primordial oscillation with a frequency proportional to $k$ proposed in 
Ref.~\cite{Planck:2018jri}. In Sec.~\ref{sec:model}, we introduce the model studied and we 
describe the comparison between its patter and the lensing contribution on the CMB spectra and 
the baryon acoustic oscillations (BAO) signal on the matter power spectrum. We describe and 
show the results from our simulations in Sec.~\ref{sec:sims} and we compare them to the 
predictions from perturbation theory in Sec.~\ref{sec:perturbation}.
Finally, we derive the constraint on the feature amplitude using the two-point correlation 
function from BOSS DR12 in \ref{sec:comparison2}. Sec.~\ref{sec:conclusion} contains our 
conclusions.

\section{Primordial oscillatory features with a Gaussian envelope} \label{sec:model}
As show in \cite{Domenech:2019cyh,Planck:2018jri}, the effects of the phenomenological lensing 
parameter $A_{\rm L}$ can be mimicked by injecting in the PPS a feature oscillating linearly in 
$k$ with a frequency and a phase matching to the one of the acoustic peaks of the CMB angular 
power spectra. In particular, it is fundamental that the oscillations have a scale-dependent 
modulation in order to reproduce $A_{\rm L} \simeq 1$. We study the following template with 
damped linear oscillations
\begin{equation} \label{eqn:template}
    {\cal P}_{\cal R}(k) = {\cal P}_{{\cal R},\,0}(k) \left[1 + {\cal A}_{\rm lin} e^{-\frac{\left(k-\mu_{\rm env}\right)^2}{2\sigma_{\rm env}^2}}\cos\left(\omega_{\rm lin}\frac{k}{k_*} + \phi_{\rm lin}\right)\right] \,.
\end{equation}
where ${\cal P}_{{\cal R},\,0}(k) = A_{\rm s} \left(k/k_*\right)^{n_{\rm s}-1}$ is the standard 
power-law PPS of the comoving curvature perturbations $\cal R$. $A_{\rm s}$ and $n_{\rm s}$ are 
the amplitude and the spectral index of the comoving curvature perturbations at a given pivot 
scale $k_* = 0.05\ {\rm Mpc}^{-1}$.
Because of the dependence of the template on five parameters and the degeneracy among them, the 
parameters for which the template mimics the effect $A_{\rm L}$ are not uniquely determined. We 
will consider as examples for the parameters to mimic $A_{\rm L} \simeq 1$ those quoted in 
\cite{Planck:2018jri}: ${\cal A}_{\rm lin} = 0.16$, $\omega_{\rm lin} = 10^{1.158} \simeq 14.4$, 
$\mu_{\rm env} = 0.2\,{\rm Mpc}^{-1}$, $\sigma_{\rm env} = 0.057\,{\rm Mpc}^{-1}$, 
and $\phi_{\rm lin} = \pi$. This choice is not however important for our considerations and 
there might other choices of parameters which also show the degeneracy of the template in 
Eq.~\eqref{eqn:template} with $A_{\rm L} \simeq 1$.

\begin{figure}
\centering
\includegraphics[width=0.48\textwidth]{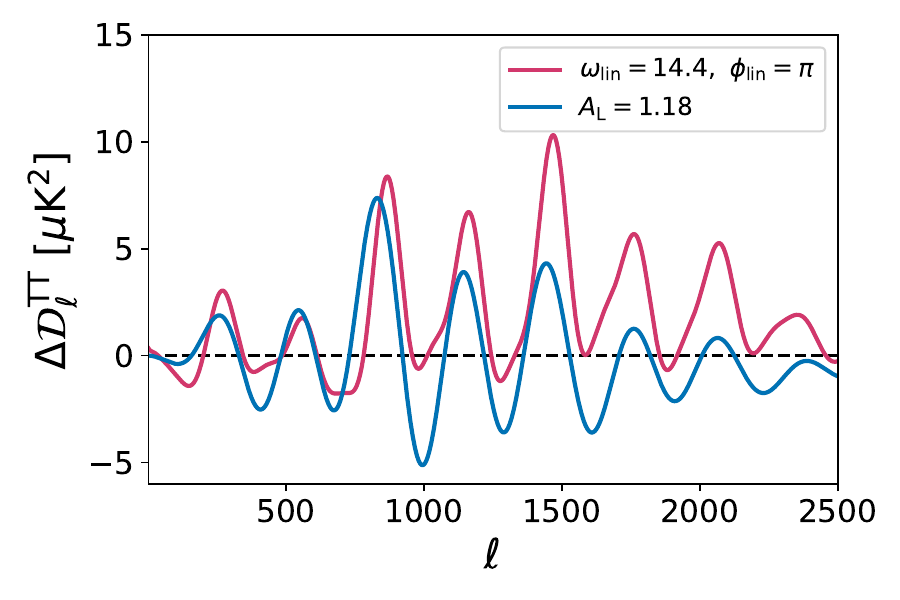}
\includegraphics[width=0.48\textwidth]{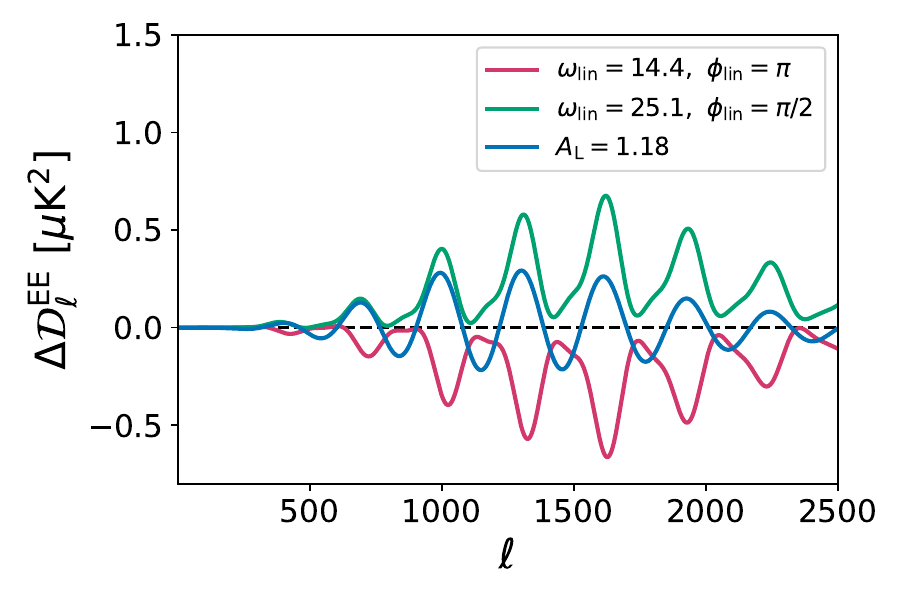}
\includegraphics[width=0.48\textwidth]{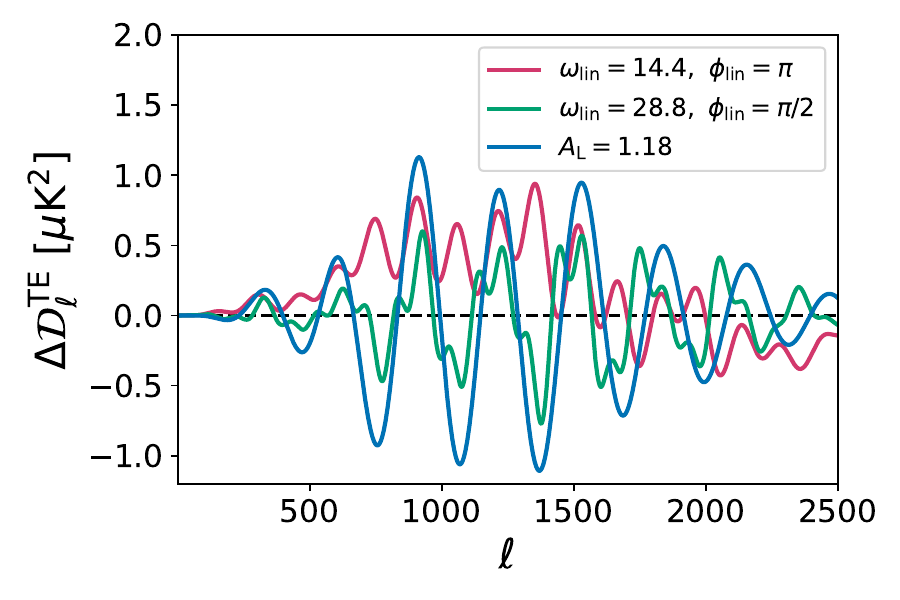}
\caption{Differences with respect to the $\Lambda$CDM CMB angular power spectrum for the case with 
$A_{\rm L} = 1.18$ (blue) and for the features template \eqref{eqn:template}.} 
\label{fig:comparison}
\end{figure}

This particular template matches reasonably well the residuals between the CMB temperature 
angular power spectrum and the $\Lambda$CDM bestfit, but does not reproduce the residuals 
in polarization and temperature-polarization cross-correlation, as already emphasized in 
\cite{Planck:2018jri}. In Fig.~\ref{fig:comparison}, we show the comparison between the feature 
best-fit and the featureless power spectrum with $A_{\rm L} = 1.18$. While 
Eq.~\eqref{eqn:template} with the best-fit parameters above can recover the residuals for the 
case with $A_{\rm L} = 1.18$ for the CMB temperature, it does not for the E-mode polarization 
and the temperature-polarization cross spectra unless we change both the frequency and the 
phase.

\begin{figure}
\centering
\includegraphics[width=0.48\textwidth]{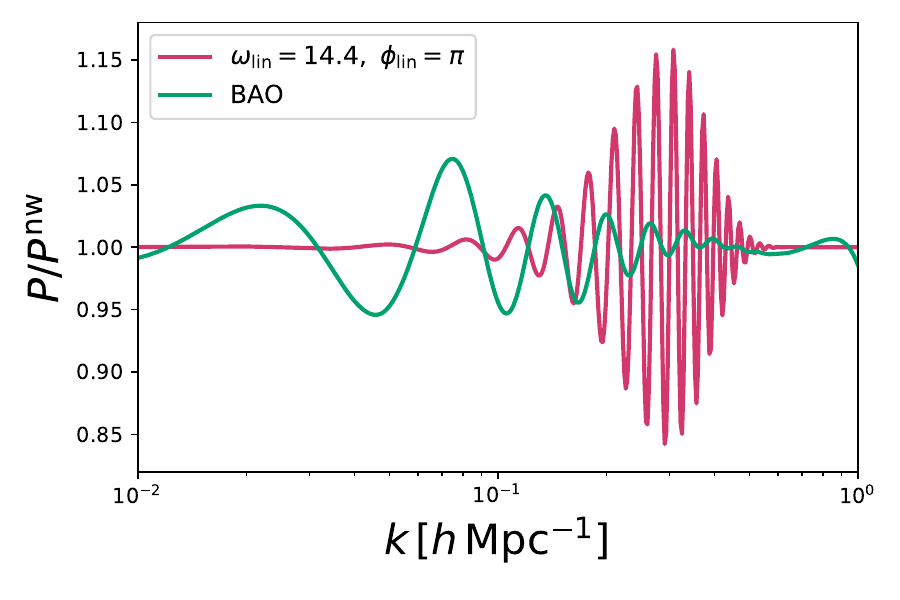}
\includegraphics[width=0.48\textwidth]{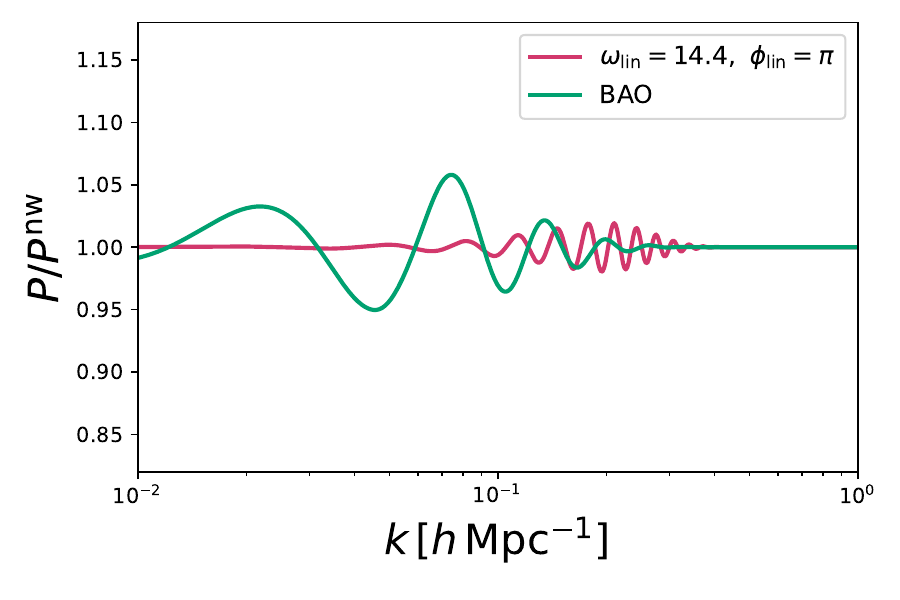}
\caption{Comparison between the BAO signal (green line) and the template \eqref{eqn:template} 
with $A_{\rm lin} = 0.16$, $\log_{10} \omega_{\rm lin} = 1.158$, $\phi_{\rm lin} = \pi$ (magenta 
line). {\em Left panel:} ratio of the linear matter power spectrum at $z = 0$. {\em Right panel:} 
ratio of the non-linear matter power spectrum at $z = 0$.}
\label{fig:comparison_BAO}
\end{figure}

In Fig.~\ref{fig:comparison_BAO}, we compare the template with damped linear oscillations with 
BAO feature at redshift $z = 0$. The two signals have different frequency, phase and they peak 
at different wave-number.

Finally, the CMB lensing power spectrum is not very sensitive to this class of primordial 
features with high frequencies \cite{Chandra:2021ydm}. 
The power spectrum of the lensing potential is an integrated quantity where a large range of wavenumbers 
contribute to each multipole; as a result, high frequency oscillations as those considered here are efficiently smoothed in CMB lensing compared to CMB temperature and polarization spectra.

\section{Cosmological simulations with primordial features} \label{sec:sims}
In order to study the effect of such signal imprinted on the matter power spectrum, we have run 
a modified version of the publicly available code 
{\tt L-PICOLA}\footnote{\href{https://github.com/CullanHowlett/l-picola}{https://github.com/CullanHowlett/l-picola}} \cite{Tassev:2013pn,Howlett:2015hfa,Winther:2017jof} 
to produce a total of nine simulations with different cosmological parameters. 
We have fixed the standard cosmological parameters to $h = 0.6736$, $\Omega_{\rm m} = 0.31377$, 
$\Omega_{\rm b} = 0.04930$, and $\sigma_8 = 0.8107$ according to Ref.~\cite{Planck:2018vyg}.
Initially, we start with the fiducial feature parameters described in \cite{Planck:2018jri} 
(hereafter M1) and the corresponding featureless case. Subsequently, we have studied the effect of varying one of the feature parameters at the time: 
M2 with ${\cal A}_{\rm lin} \to {\cal A}_{\rm lin}/2 = 0.08$, 
M3 with ${\cal A}_{\rm lin} \to 2{\cal A}_{\rm lin} = 0.32$, 
M4 with $\mu_{\rm env} \to \mu_{\rm env}/2 = 0.1\,{\rm Mpc}^{-1}$, 
M5 with $\mu_{\rm env} \to 2\mu_{\rm env} = 0.4\,{\rm Mpc}^{-1}$, 
M6 with $\sigma_{\rm env} \to \sigma_{\rm env}/2 = 0.0285\,{\rm Mpc}^{-1}$, 
M7 with $\sigma_{\rm env} \to 2\sigma_{\rm env} = 0.114\,{\rm Mpc}^{-1}$, 
M8 with $\log_{10} \omega_{\rm lin} \to 0.8 \log_{10} \omega_{\rm lin} = 0.926$, 
and M9 with $\log_{10} \omega_{\rm lin} \to 1.2 \log_{10} \omega_{\rm lin} = 1.39$.

Each simulation has $1024^3$ dark matter particles with a comoving box with side length of 
$1024\,h^{-1}\,{\rm Mpc}$ evolved with 30 time steps\footnote{We found consistent and robust 
results by running smaller simulations with $512^3$ particles and a box with side of 
$512\,h^{-1}\,{\rm Mpc}$.}. 
The initial conditions are produced using second-order Lagrangian Perturbation Theory, with the 
{\tt 2LPTic} code~\cite{Crocce:2006ve}, at redshift $z = 9$, with input the linear 
matter power spectra were computed with a modified version of the publicly available code 
{\tt CAMB}\footnote{\href{https://github.com/cmbant/CAMB}{https://github.com/cmbant/CAMB}} \cite{Lewis:1999bs}, as in \cite{Ballardini:2019tuc}.

In order to minimize cosmic variance, we run pair of simulations with the same initial seed, 
inverted initial condition, and with fixed amplitude as in \cite{Ballardini:2019tuc}. 
Finally, we use observable averaged over these two simulations.
Using paired-fixed simulations, with paired phase \cite{Viel:2010bn,Pontzen:2015eoh} 
and fixed amplitude \cite{Viel:2010bn,Angulo:2016hjd}, significantly reduces the variance of the N-body 
simulations; see Ref.~\cite{Villaescusa-Navarro:2018bpd} for a comprehensive study of this procedure. 

\begin{figure}
\centering
\includegraphics[width=0.9\textwidth]{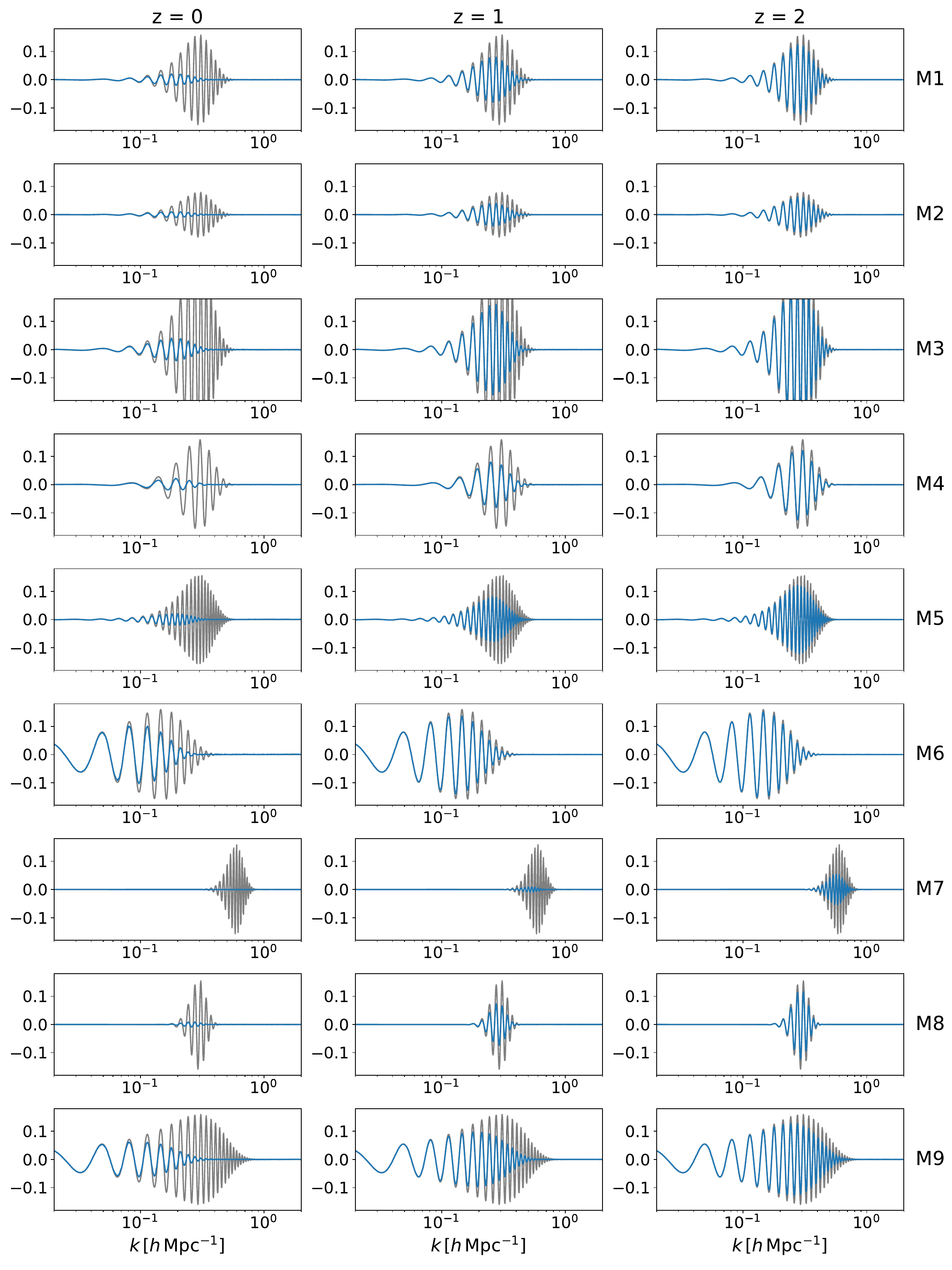}
\caption{Relative differences with respect to the $\Lambda$CDM matter power spectrum for the 
linear matter power spectrum (gray) and the nonlinear matter power spectrum obtained from the 
simulations (blue) for the template with damped linear oscillations \eqref{eqn:template} at 
redshift $z = 0,\,1,\,2$.}
\label{fig:COLA}
\end{figure}

In Fig.~\ref{fig:COLA}, we show the relative differences between the nonlinear matter power 
spectrum of the feature model \eqref{eqn:template} with respect to the featureless case for 
the feature bestfit and the eight models (from the top to the bottom) at different redshift 
$z = 0,\,1,\,2$ (from the left to the right).

\section{Perturbation theory with primordial features} \label{sec:perturbation}
Following the perturbative approach in 
Refs.~\cite{Beutler:2019ojk,Vasudevan:2019ewf}, closely connected to the BAO resummation done in 
\cite{Blas:2015qsi,Blas:2016sfa}, we start by decomposing the linear power spectrum into a smooth 
(nw) and an oscillating (w) contribution
\begin{equation} \label{eqn:split1}
    P^{\rm lin}(z,k) = G^2(z) \left[P^{\rm nw}(k) + P^{\rm w}(k)\right] \,.
\end{equation}
Here the oscillatory part describes both BAO and the primordial oscillations as
\begin{equation} \label{eqn:split2}
    P^{\rm w}(k) \equiv P^{\rm nw}\left[\delta P^{\rm w}_{\rm BAO}(k) 
    + \delta P^{\rm w}_{\rm lin}(k) + \delta P^{\rm w}_{\rm BAO}(k)\delta P^{\rm w}_{\rm lin}(k)\right]
\end{equation}
with
\begin{equation}
    \delta P^{\rm w}_{\rm lin}(k) = {\cal A}_{\rm lin} 
    e^{-\frac{\left(k-\mu_{\rm env}\right)^2}{2\sigma_{\rm env}^2}}
    \cos\left(\omega_{\rm lin}\frac{k}{k_*} + \phi_{\rm lin}\right)
\end{equation}
and $k_* = 0.05$ Mpc$^{-1}$.
Here we have factored out the time-dependence given by the growth factor $G(z)$. The cross term 
in Eq.~\eqref{eqn:split2} is subdominant since it is proportional to 
$A_{\rm BAO} \cdot {\cal A}_{\rm lin} \simeq 0.01$ and we neglect it.

The IR resummed power spectrum at leading order (LO) is given by
\begin{equation} \label{eqn:IR_LO_1}
    P^{\rm IR\, res,\, LO}(z,k) = G^2(z) \left[P^{\rm nw}(k) 
    + e^{-k^2 G^2(z) \Sigma^2} P^{\rm w}(k)\right]
\end{equation}
where $P^{\rm nw}(k)$ corresponds to the smooth part of the linear power spectrum and $P^{\rm w}(k)$ is multiplied by the exponential damping of the oscillatory part due to the effect of IR enhanced loop contributions, exactly as for BAO \cite{Blas:2016sfa}. For BAO, the damping factor corresponds to
\begin{equation} \label{eqn:sigma}
    \Sigma_{\rm BAO}^2(k_{\rm S}) \equiv \int_0^{k_{\rm S}}\frac{{\rm d}q}{6\pi^2} P^{\rm nw}(q) \left[1-j_0\left(q r_{\rm s}\right)+2j_2\left(q r_{\rm s}\right)\right]
\end{equation}
where $j_n$ are spherical Bessel functions and $k_{\rm S}$ is the separation scale of long and 
short modes, that has been introduced in order to treat the perturbative expansion in the two 
regimes separately. The dependence of the spectrum from $k_{\rm S}$ can be connected with an 
estimate of the perturbative uncertainty. For this reason and since IR expansions are valid for 
$q \ll k$, we assumed $k_{\rm S} = \epsilon k$ with $\epsilon \in [0.3,0.7]$ as in 
Ref.~\cite{Vasudevan:2019ewf}. $r_{\rm s} \simeq 147\,{\rm Mpc}$ is the scale setting the period of 
the BAO \cite{Planck:2018vyg}. For the primordial feature, the damping factor 
\eqref{eqn:sigma} depends also from the frequency $\omega_{\rm lin}$ as
\begin{equation} \label{eqn:sigma_lin}
    \Sigma_{\rm lin}^2(\omega_{\rm lin},k_{\rm S}) \equiv 
    \int_0^{k_{\rm S}}\frac{{\rm d}q}{6\pi^2} P^{\rm nw}(q) \left[1-j_0\left(q \frac{\omega_{\rm lin}}{k_*}\right)+2j_2\left(q \frac{\omega_{\rm lin}}{k_*}\right)\right] \,.
\end{equation}
We can rewrite Eq.~\eqref{eqn:IR_LO_1} as
\begin{align} \label{eqn:IR_LO_2}
    P^{\rm IR\, res,\, LO}(z,k) = G^2(z) P^{\rm nw}(k)
    &\left[1 + e^{-k^2 G^2(z) \Sigma^2_{\rm BAO}} \delta P^{\rm w}_{\rm BAO}(k)\right. \notag\\
    &\left.+e^{-k^2 G^2(z) \Sigma^2_{\rm lin}}\delta P^{\rm w}_{\rm lin}(k)\right] \,.
\end{align}

At next-to-leading order (NLO), the IR resummed power spectrum can be written in the form
\begin{align} \label{eqn:bao_NLO}
    P^{\rm IR\, res,\, LO+NLO}(z,k) = G^2(z) P^{\rm nw}(k) 
    &\left\{1 + \left[1+k^2 G^2(z) \Sigma^2_{\rm BAO}\right]e^{-k^2 G^2(z) \Sigma^2_{\rm BAO}} \delta P^{\rm w}_{\rm BAO}(k)\right. \notag\\
    &\left.+\left[1+k^2 G^2(z) \Sigma^2_{\rm lin}\right]e^{-k^2 G^2(z) \Sigma^2_{\rm lin}}\delta P^{\rm w}_{\rm lin}(k)\right\} \notag\\
    &+ G^4(z) P^{\rm 1-loop}\left[ P^{\rm IR\, res,\, LO}(k) \right]
\end{align}
where we neglected the leading contribution to the NLO coming from 2-loops being numerically 
small (see Ref.~\cite{Blas:2016sfa} for the full expression at NLO for the resummed featureless 
matter power spectrum). We also neglect contribution to the power spectrum coming from 
the bispectrum that might became relevant for higher frequency compared to the one studied here. 
$P^{\rm 1-loop}$ is the standard one-loop result, but computed with the LO IR resummed power 
spectrum. In practice, one can use the usual expression $P^{\rm 1-loop} = P_{22} + 2 P_{13}$, 
however evaluating the loop integrals $P_{22}$ and $P_{13}$ with the input spectrum $P^{\rm IR\, 
res,\, LO}$ instead of the linear spectrum \cite{Blas:2015qsi}. 
Note that we do not consider for any correction due to the Gaussian envelop just considering the 
linear oscillatory pattern on top of the matter power spectrum as it happens for BAO \cite{Blas:2016sfa}.

\subsection{Comparison with cosmological simulations} \label{sec:comparison1}
\begin{figure}
\centering
\includegraphics[width=0.48\textwidth]{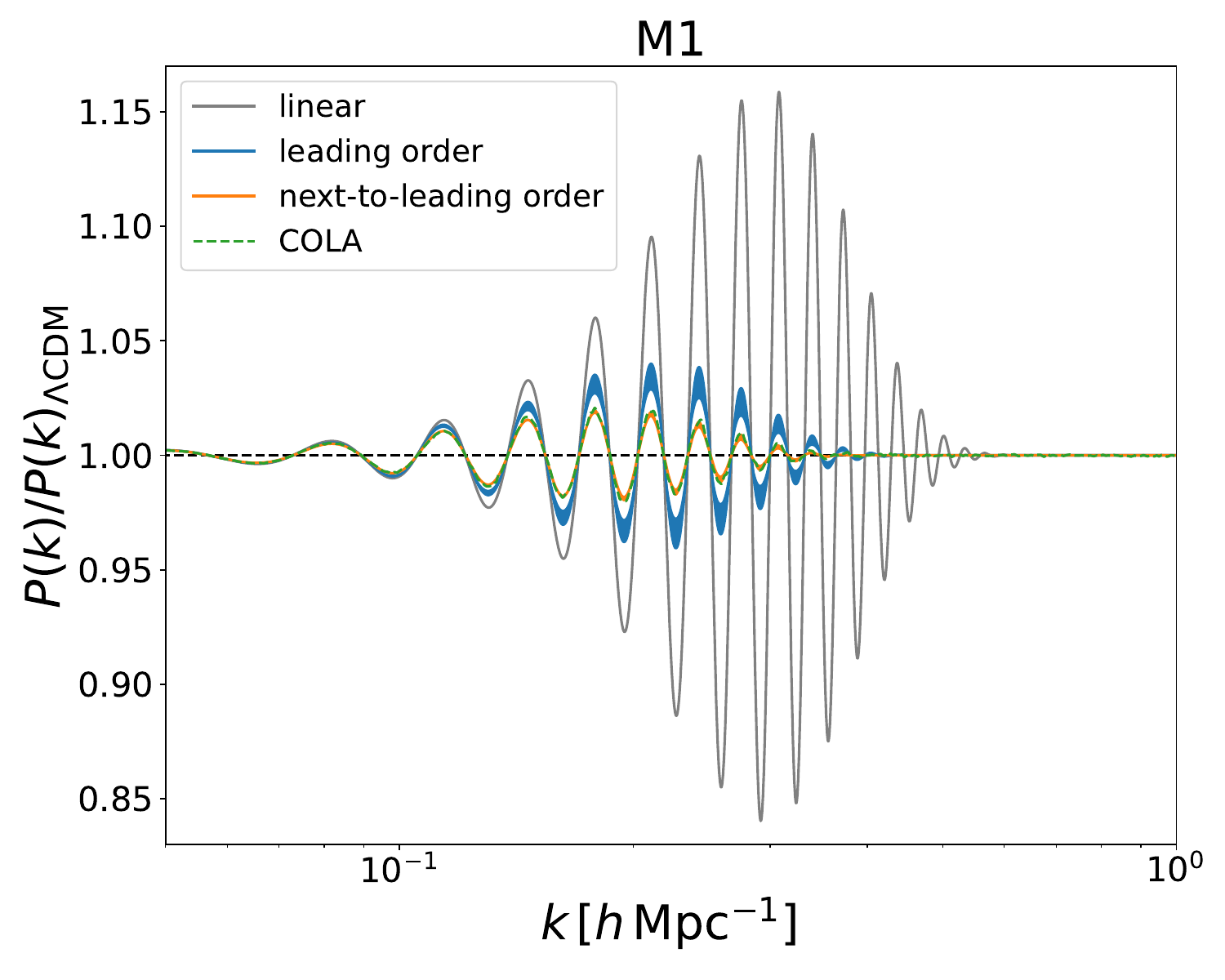}
\includegraphics[width=0.48\textwidth]{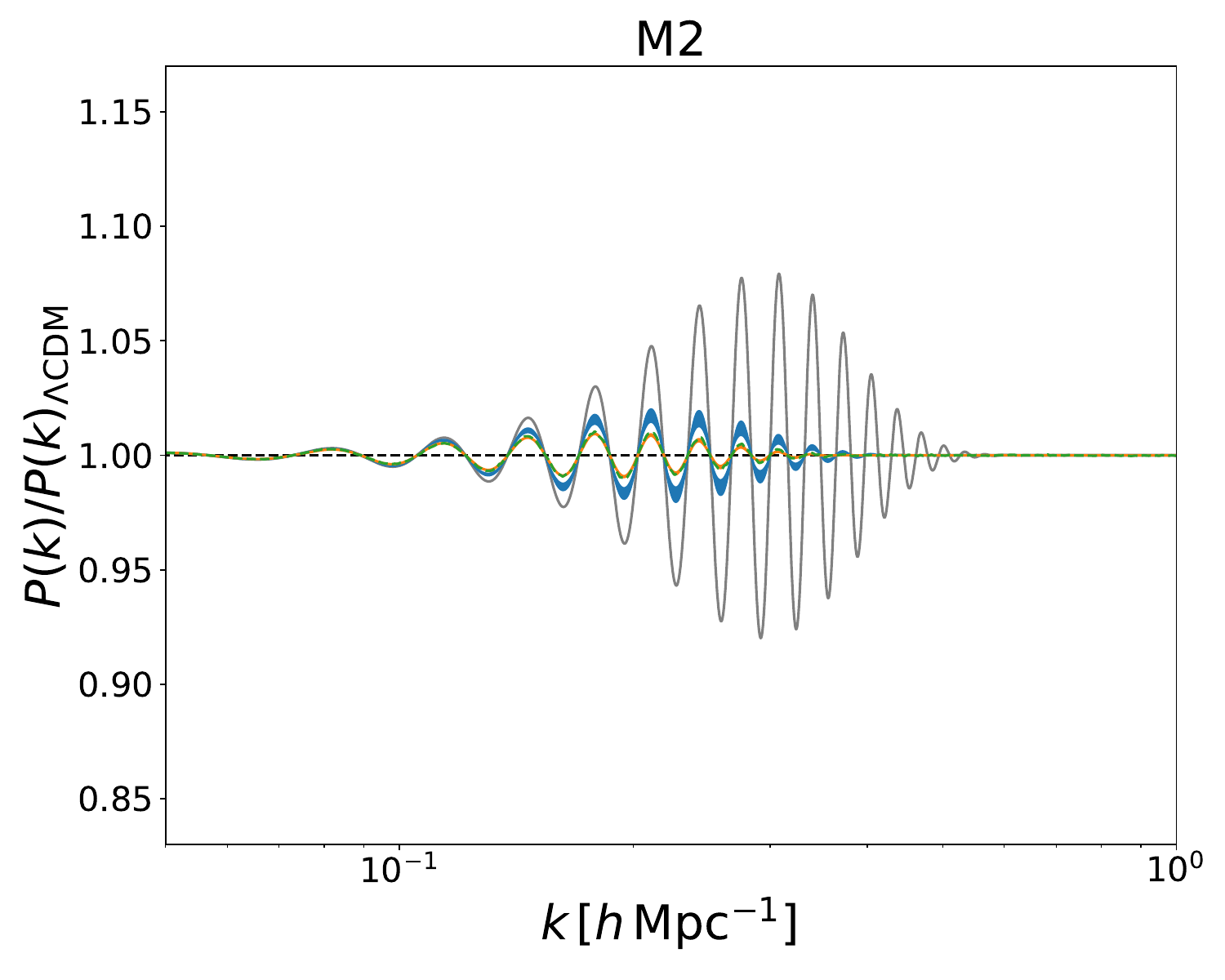}
\includegraphics[width=0.48\textwidth]{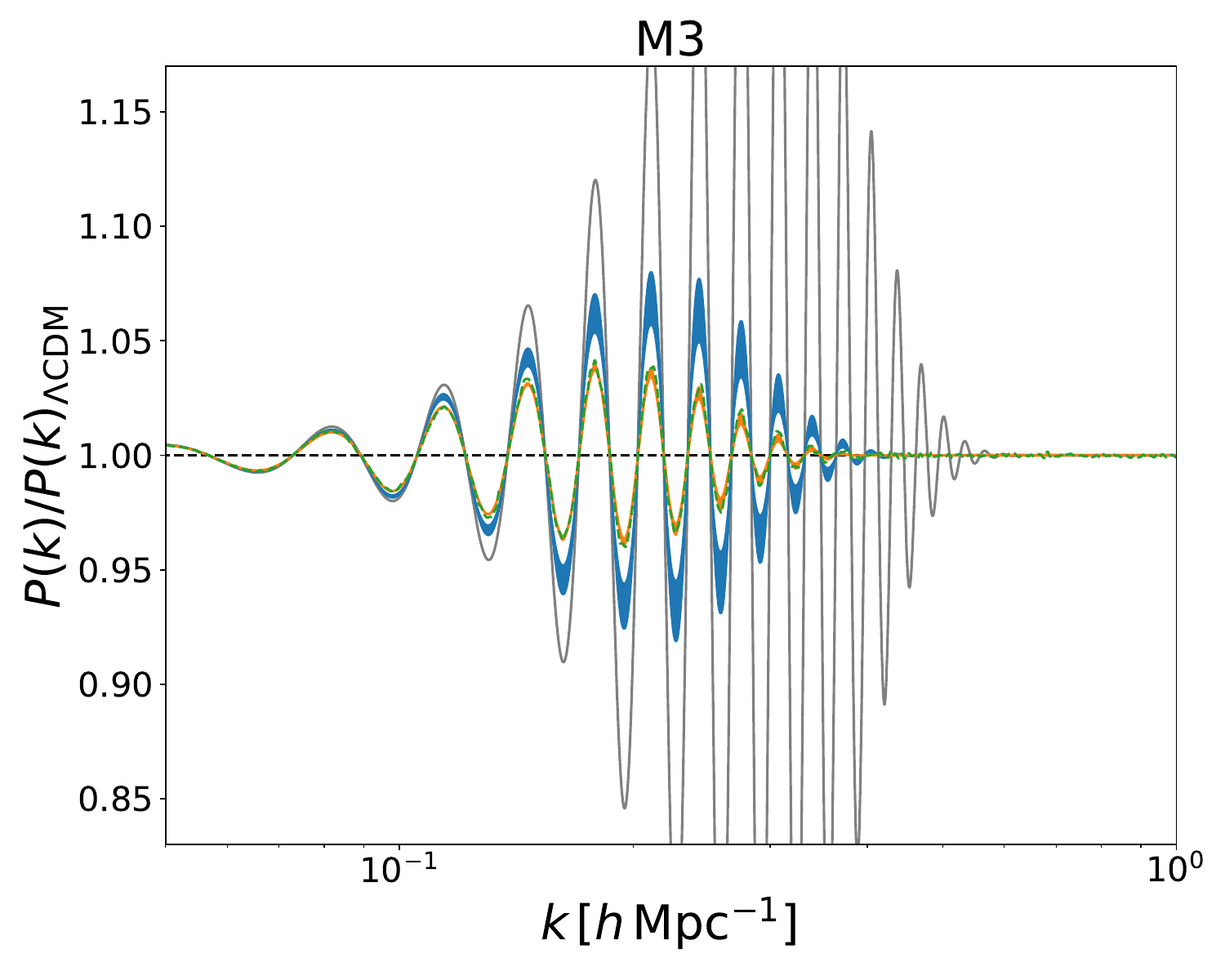}
\caption{Ratio of IR resummed matter power spectrum at LO (blue) and NLO (orange) obtained for 
the damped linear oscillations to the one obtained with a power-law PPS at redshift $z = 0$ 
varying the IR separation scale $k_{\rm S} = \epsilon k$ with $\epsilon \in [0.3,0.7]$. Also 
shown is the linear result (gray). We show the results obtained from simulations (green dashed) 
for the best-fit parameters M1 (top left panel) and we change the value of the feature amplitude 
to ${\cal A}_{\rm lin} = 0.08$ (top right panel) and ${\cal A}_{\rm lin} = 0.32$ (bottom 
panel).}
\label{fig:SPT_1}
\end{figure}

\begin{figure}
\centering
\includegraphics[width=0.48\textwidth]{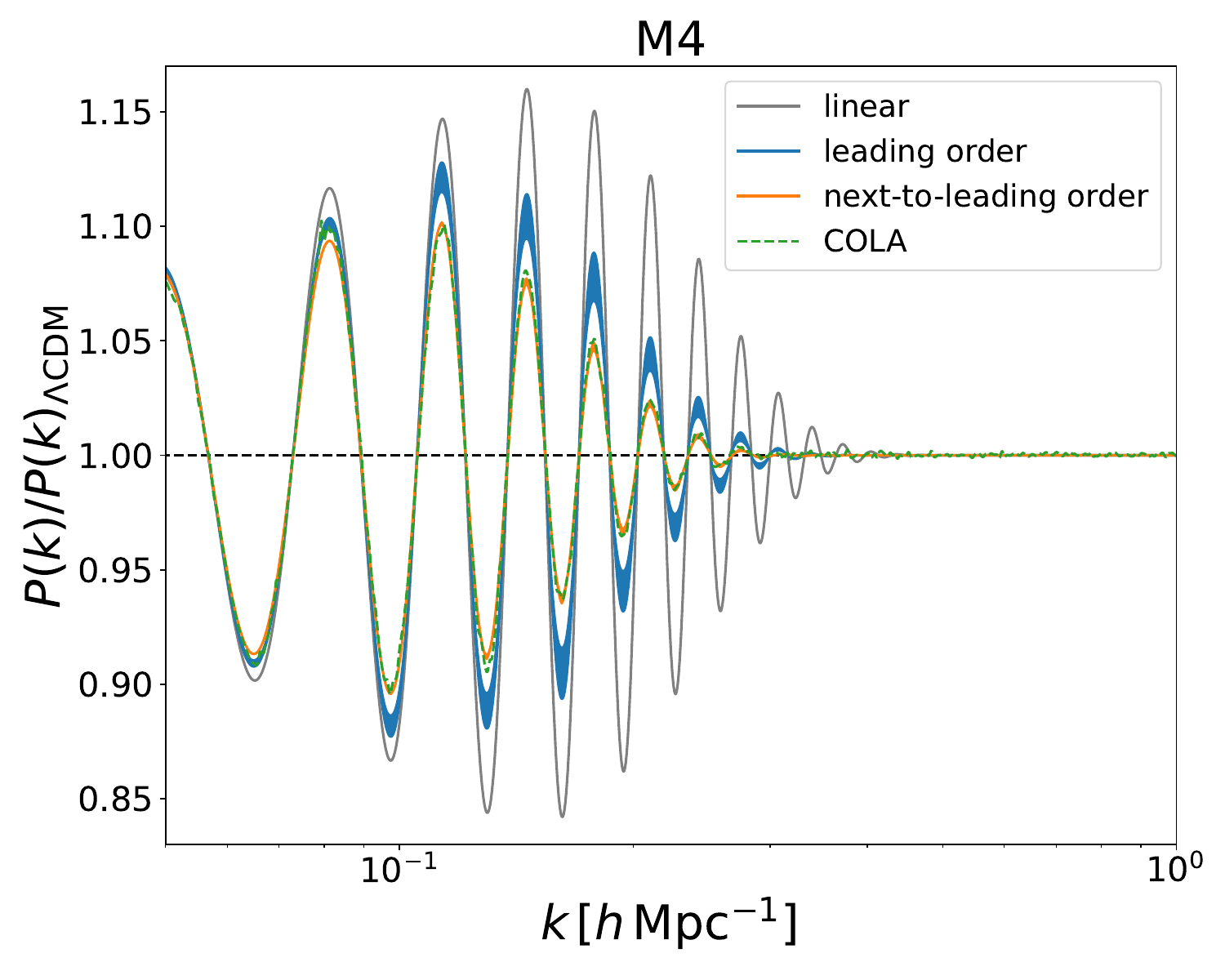}
\includegraphics[width=0.48\textwidth]{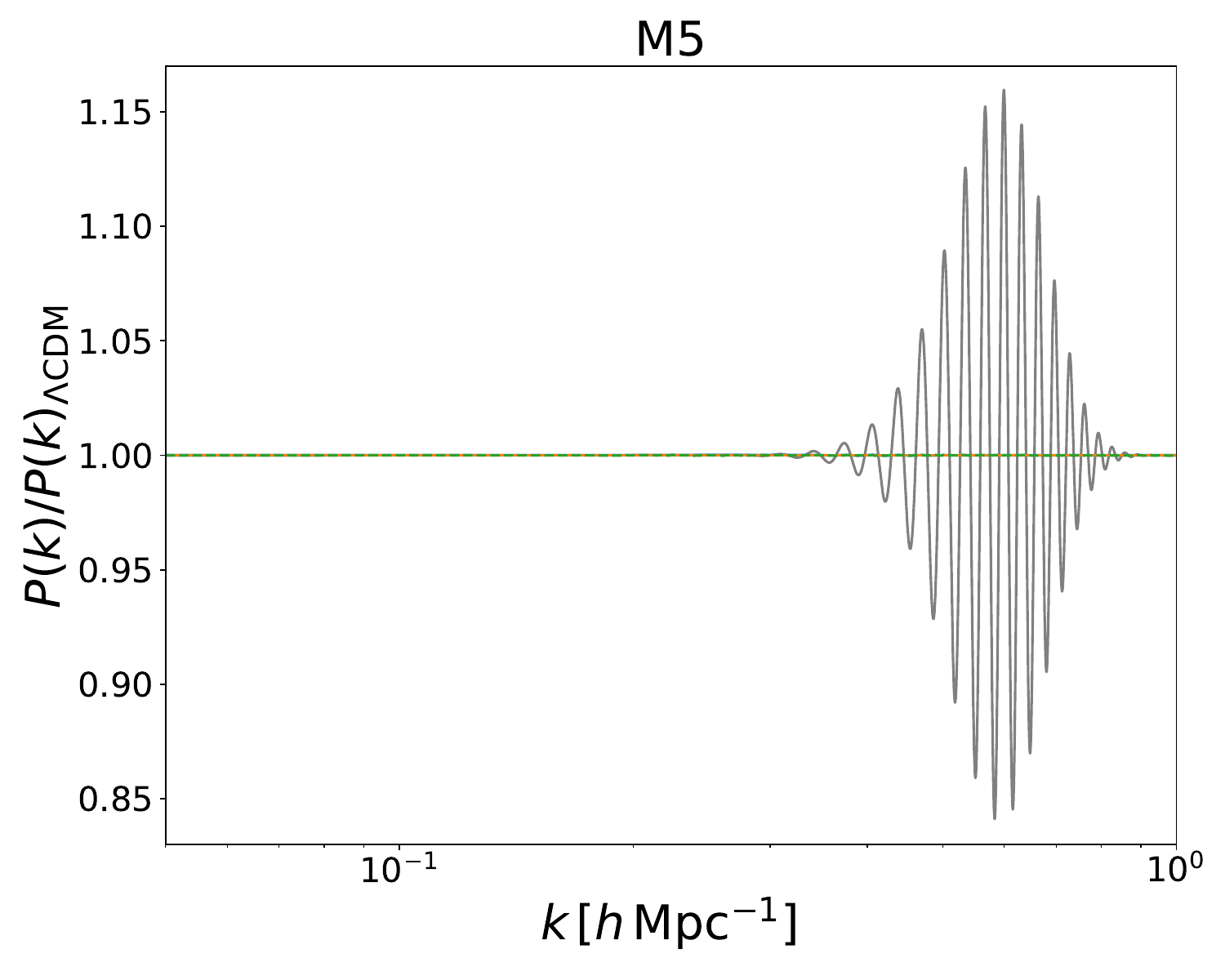}
\caption{As Fig.~\ref{fig:SPT_1} for $\mu_{\rm env} = 0.1\ {\rm Mpc}^{-1}$ (left panel) and for 
$\mu_{\rm env} = 0.4\ {\rm Mpc}^{-1}$ (right panel).}
\label{fig:SPT_2}
\end{figure}

\begin{figure}
\centering
\includegraphics[width=0.48\textwidth]{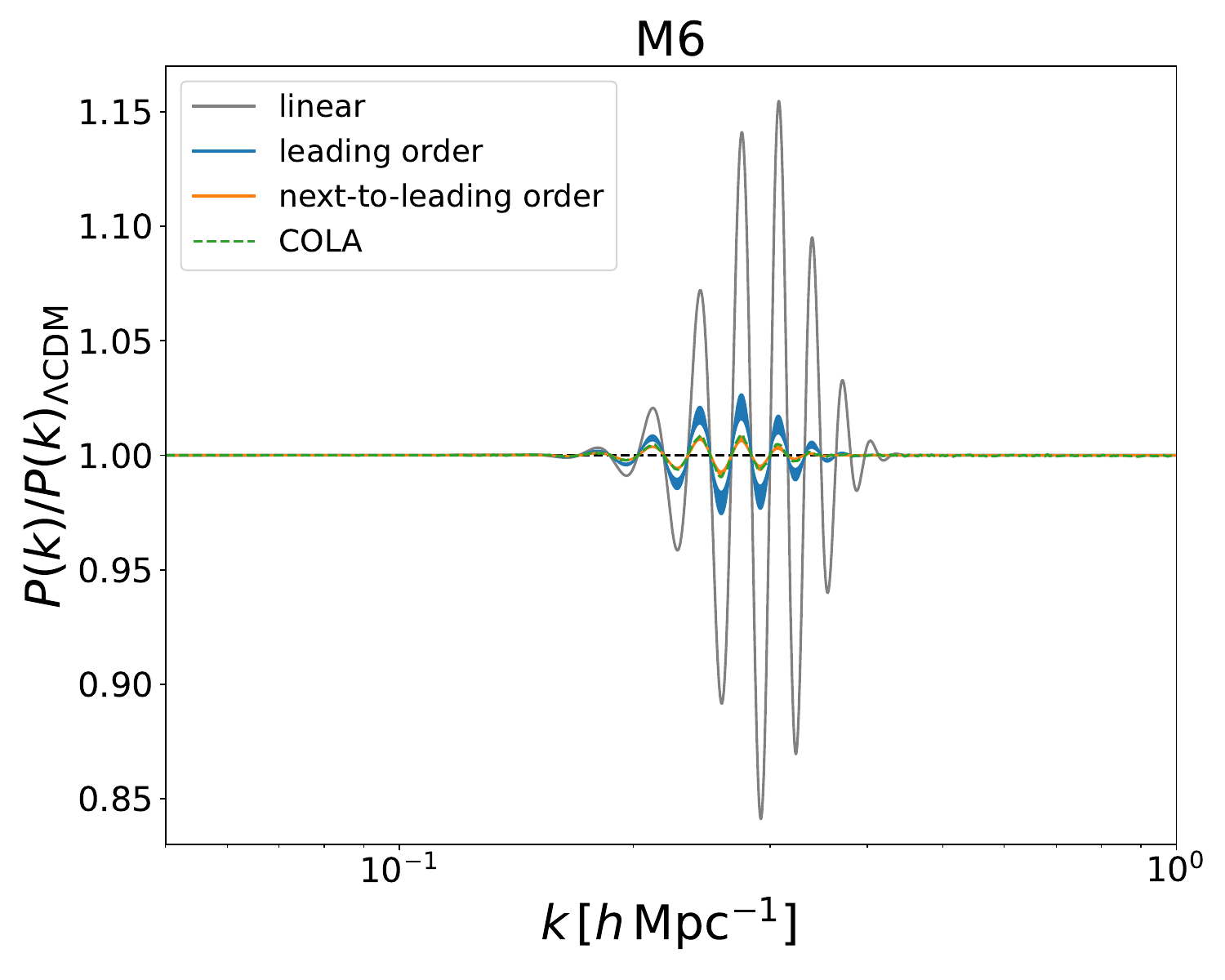}
\includegraphics[width=0.48\textwidth]{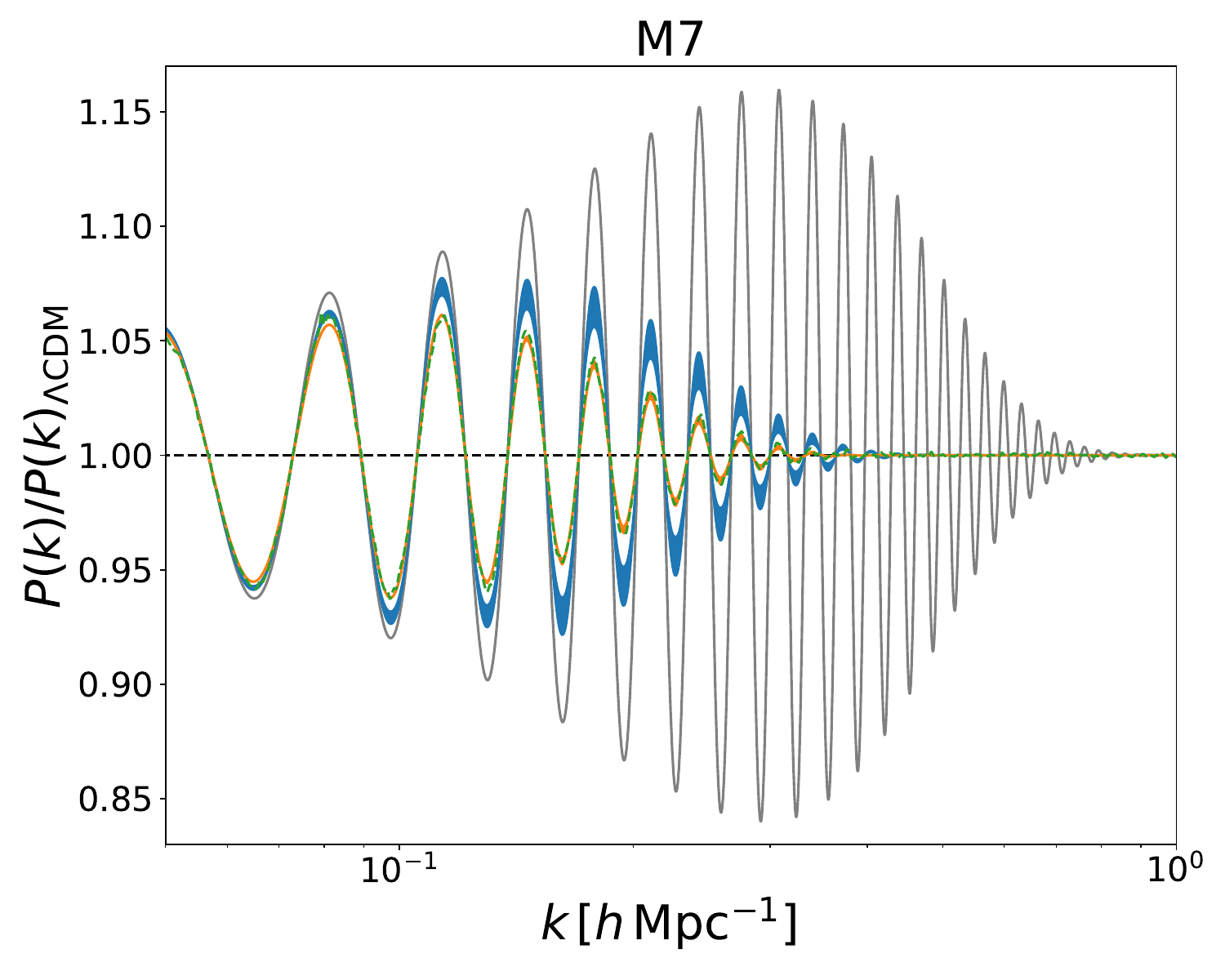}
\caption{As Fig.~\ref{fig:SPT_1} for $\sigma_{\rm env} = 0.0285\ {\rm Mpc}^{-1}$ (left panel) and 
for $\sigma_{\rm env} = 0.114\ {\rm Mpc}^{-1}$ (right panel).}
\label{fig:SPT_3}
\end{figure}

\begin{figure}
\centering
\includegraphics[width=0.48\textwidth]{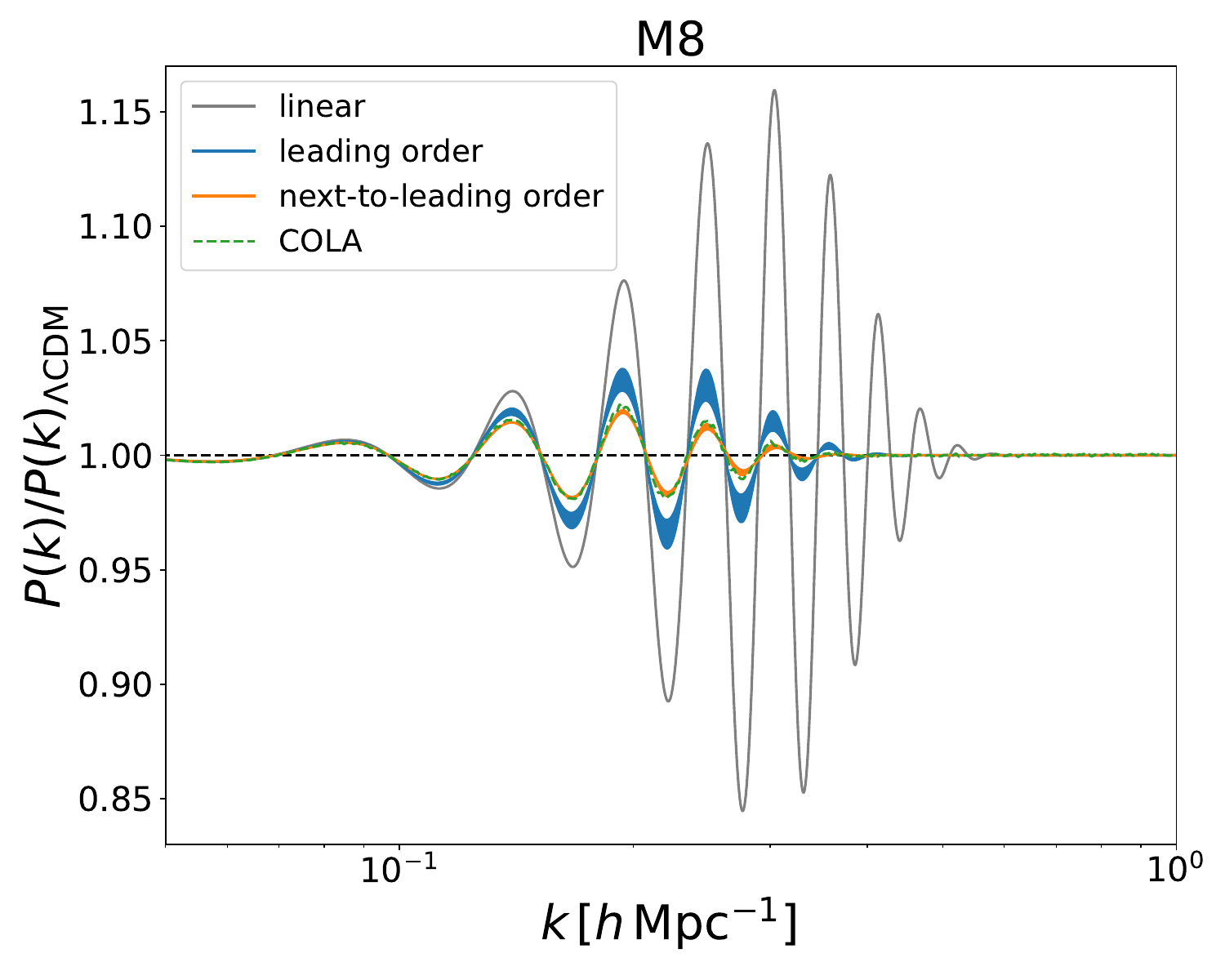}
\includegraphics[width=0.48\textwidth]{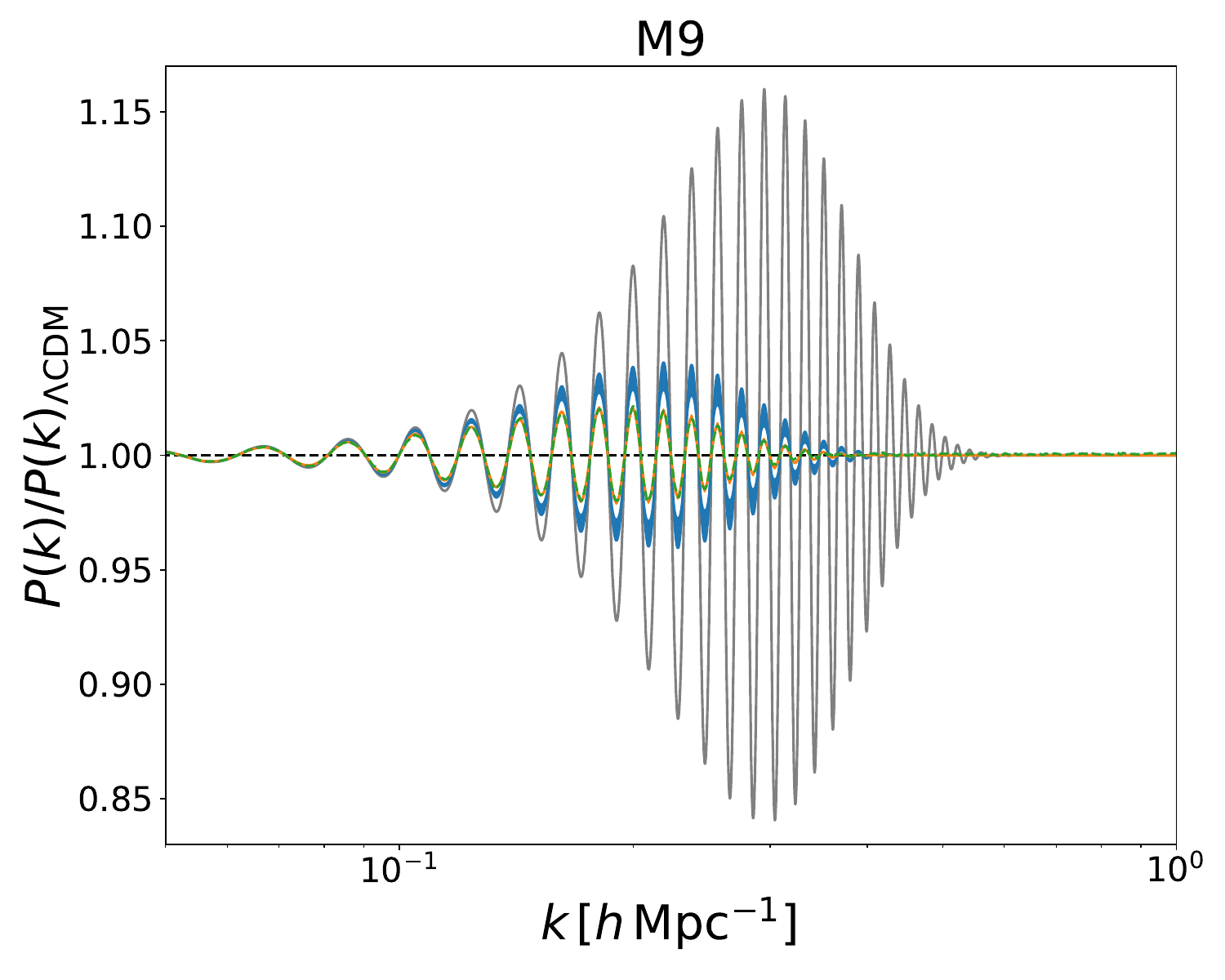}
\caption{As Fig.~\ref{fig:SPT_1} for $\log_{10} \omega_{\rm lin} = 0.926$ (top panel) and 
for $\log_{10} \omega_{\rm lin} = 1.39$ (bottom panel).}
\label{fig:SPT_4}
\end{figure}
In Figs.~\ref{fig:SPT_1}-\ref{fig:SPT_2}-\ref{fig:SPT_3}-\ref{fig:SPT_4}, we show the ratio 
between the matter power spectrum with damped primordial oscillations and the one with power-law 
PPS calculated at redshift $z = 0$ for the results both at LO and NLO. In particular, we vary the 
amplitude ${\cal A}_{\rm lin}$ in Fig.~\ref{fig:SPT_1}, the mean of the Gaussian envelope 
$\mu_{\rm env}$ in Fig.~\ref{fig:SPT_2}, the dispersion of the Gaussian envelope 
$\sigma_{\rm env}$ in Fig.~\ref{fig:SPT_3}, and the feature frequency $\log_{10} \omega_{\rm lin}$ 
in Fig.~\ref{fig:SPT_4}. 

\begin{figure}
\centering
\includegraphics[width=0.9\textwidth]{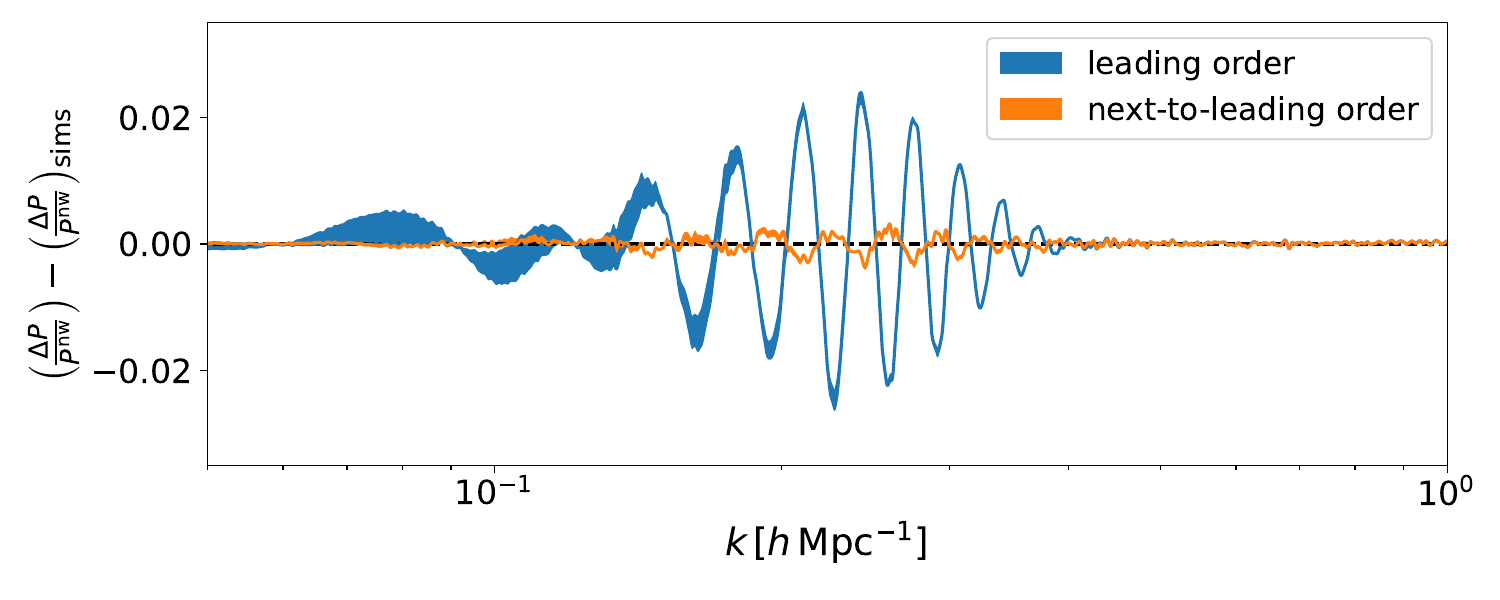}
\caption{Comparison of the primordial oscillatory component of the matter power spectrum 
as predicted by LO (blue) and NLO (orange) with respect to the result from N-body simulations 
varying the IR separation scale $k_{\rm S} = \epsilon k$ with $\epsilon \in [0.3,0.7]$.}
\label{fig:diff}
\end{figure}

One can observe that the agreement between simulations and perturbation theory predictions is 
considerably improved going from LO to NLO.
Furthermore, the dependence on the separation scale 
$k_{\rm S}$ is reduced (reducing the theoretical perturbative uncertainties).
We find for the Fourier matter power spectrum at $z = 0$ differences less than 3\% for the LO and 
less than 0.3\% including the NLO with respect to our DM-only simulations, see 
Fig.~\ref{fig:diff}.
These differences and the predictions from perturbation theory are robust to the 
variation of the primordial feature parameters ${\cal A}_{\rm lin}$, $\mu_{\rm env}$, 
$\sigma_{\rm env}$, and $\log_{10} \omega_{\rm lin}$.

Moving the position of the feature toward smaller scales, i.e. 
$\mu_{\rm env} > 0.2\,{\rm Mpc}^{-1}$, we see that the primordial oscillations are completely 
damped at low redshift, see Figs.~\ref{fig:COLA}-\ref{fig:SPT_2}. Analogously when we reduce 
the size of the Gaussian envelope, i.e. $\sigma_{\rm env} < 0.057\,{\rm Mpc}^{-1}$, see 
Figs.~\ref{fig:COLA}-\ref{fig:SPT_3}.

For the best-fit parameters M1, the amplitude of the oscillations is reduced by a factor 8 at 
$z = 0$ and it is reduced by half at $z = 1$. This effect is increased or reduced by varying 
the parameters of the Gaussian envelope with respect to their fiducial values. 
Moreover, all the primordial oscillations at $k > 0.3\,h\,{\rm Mpc}^{-1}$ and $z = 0$ are washed away 
by nonlinearities. This result highlight the importance and the need of high redshift clustering 
measurements, i.e. $z \gtrsim 1$, in order to study these specific features on the matter 
power spectrum.

\section{Comparison with current LSS data} \label{sec:comparison2}
We apply the methodology developed in Ref.~\cite{Ballardini:2022wzu} and already applied to the 
case of linear undamped oscillations to the current model with the publicly available library 
{\tt CosmoBolognaLib}\footnote{\href{https://gitlab.com/federicomarulli/CosmoBolognaLib}{https://gitlab.com/federicomarulli/CosmoBolognaLib}} \cite{Marulli:2015jil}. 
We consider the combination of two non-overlapping redshift bins, neglecting their correlation, 
covering $0.2 < z < 0.5$ and $0.5 < z < 0.75$ using the galaxy two-point correlation function 
(2PCF) from the Sloan Digital Sky Survey III Baryon Oscillation Spectroscopic Survey Data Release 
12 (BOSS DR12) \cite{BOSS:2016wmc,BOSS:2016apd}. 

The templates for the anisotropic 2PCF monopole and quadrupole are built starting from 
the non-linear galaxy power spectrum in redshift space \cite{Kaiser:1987qv,Fisher:1993pz,Xu:2012hg,Xu:2012fw}
\begin{equation}
    P(z,k,\mu) = \left[\frac{1+\beta\mu^2R(k,\Sigma_r)}{1+k^2\mu^2\Sigma_s^2/2}\right]^2
    P^{\rm IR res,\, LO}(z,k,\mu) \label{eqn:Pk_ani} \,.
\end{equation}
Following the BOSS DR12 2PCF analysis \cite{BOSS:2016apd}, we fix the streaming scale at 
$\Sigma_s = 4\,h^{-1}\,{\rm Mpc}$, the radial and transverse components of the standard Gaussian damping 
of BAO at $\Sigma_\parallel = 4\,h^{-1}\,{\rm Mpc}$, and $\Sigma_\perp = 2.5\,h^{-1}\,{\rm Mpc}$ for 
post-reconstruction results, where $\Sigma_{\rm BAO}^2 = \mu^2\Sigma_\parallel^2+(1-\mu^2)\Sigma_\perp^2$.
$R(k,\,\Sigma_r) = 1 - e^{-k^2\Sigma_r^2/2}$ is the smoothing applied in reconstruction and 
$\Sigma_r = 15\,h^{-1}\,{\rm Mpc}$ is the smoothing scale used when deriving the displacement field 
\cite{Seo:2015eyw}. 
Given Eq.~\eqref{eqn:Pk_ani}, we define the multipole moments
\begin{equation}
    P_\ell(k,\mu) = \frac{2\ell+1}{2}\int_{-1}^{+1}{\rm d}\mu P(k,\mu)L_\ell(\mu)
\end{equation}
where $L_\ell(\mu)$ are Legendre polynomials. These are transformed to 2PCF multipole 
\begin{equation}
    \xi_\ell(s) = \frac{i^\ell}{2\pi^2} \int {\rm d}k\,k^2 P_\ell(k,\mu)j_\ell(ks)
\end{equation}
where $j_\ell(ks)$ is the $\ell$-th order spherical Bessel function. We then use
\begin{equation}
    \xi(s,\,\mu) = \sum_{\ell=0,2} \xi_\ell(s)L_\ell(\mu) \,,
\end{equation}
and we take averages over $\mu$ window to create the template
\begin{equation}
    \xi_\ell(s,\alpha_\perp,\alpha_\parallel) = \int_{-1}^{1}{\rm d}\mu P_\ell(\mu')\xi(s',\mu')
\end{equation}
where $\mu_{\rm true} = \mu\alpha_\parallel/\sqrt{\mu^2\alpha_\parallel^2+(1-\mu^2)\alpha^2_\perp}$, 
$s_{\rm true} = s\sqrt{\mu^2\alpha_\parallel^2+(1-\mu^2)\alpha^2_\perp}$.
Finally, we fit to the data using the following model for the monopole and quadrupole 
of the 2PCF
\begin{align}
&\xi_0(s) = B_0\xi_0(s, \alpha_{\perp}, \alpha_{\parallel})+A_0^0+\frac{A_0^1}{s}+\frac{A_0^2}{s^2}\, ,\\
&\xi_2(s) = \frac{5}{2}\left[B_2\xi_2(s, \alpha_{\perp}, \alpha_{\parallel})-B_0\xi_0(s, \alpha_{\perp}, \alpha_{\parallel})\right]+A_2^0+\frac{A_2^1}{s}+\frac{A_2^2}{s^2}\,.
\end{align}

\begin{figure}
\centering
\includegraphics[height=5cm]{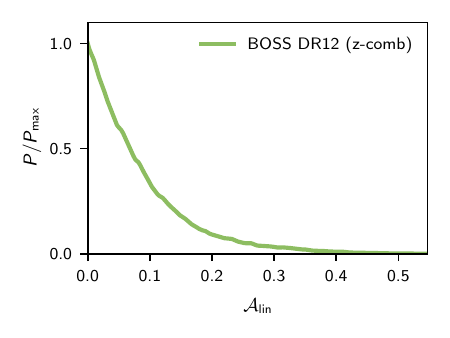}
\includegraphics[height=5.2cm]{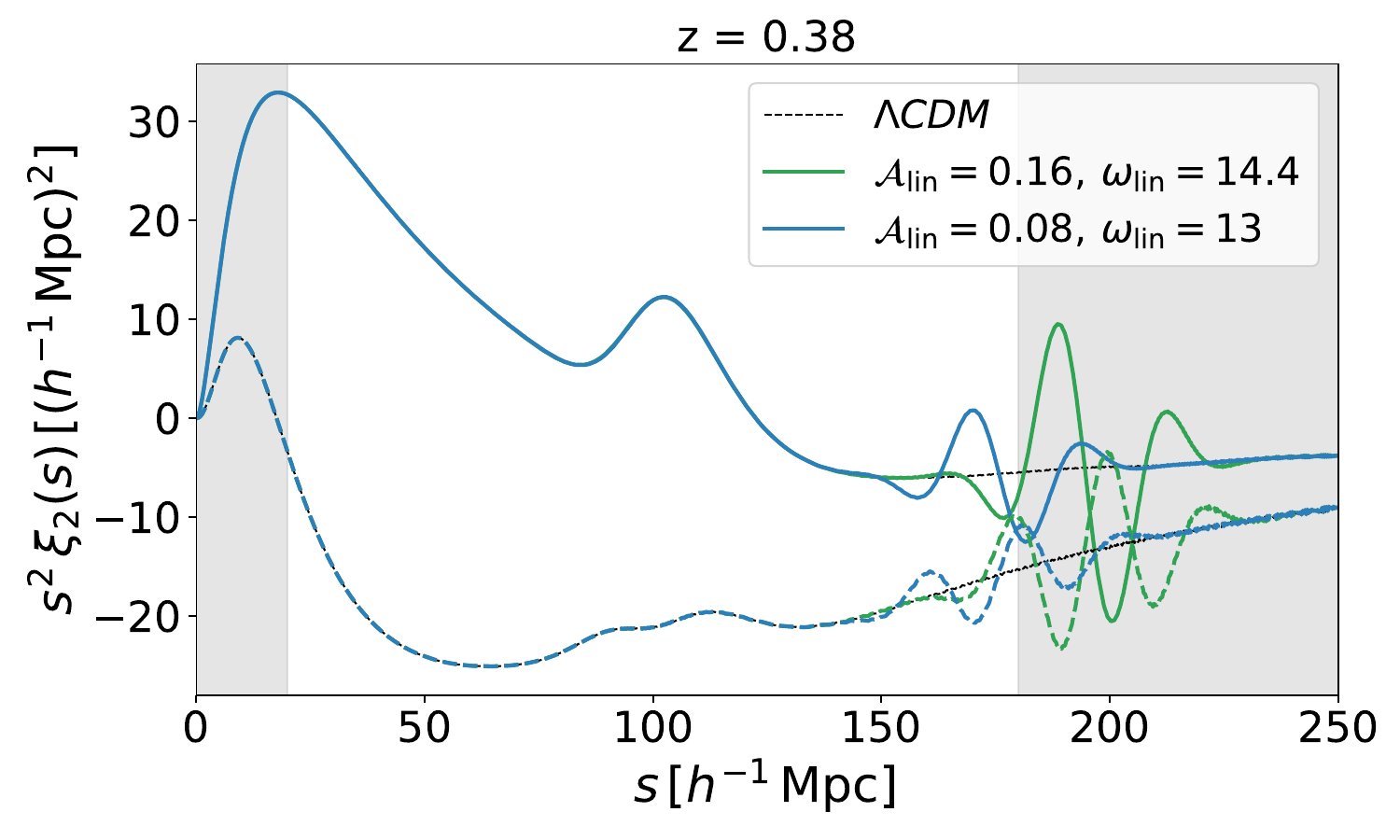}
\caption{{\em Left panel}: marginalized posterior distribution of the amplitude of the feature 
obtained with the BOSS DR12 combined redshift bin. {\em Right panel}: nonlinear CDM 2PCF monopole 
(solid) and quadrupole (dashed) computed at $z = 0.38$ (central redshift of the first 
BOSS DR12 bin) for M1 (green) and a case with half amplitude ${\cal A}_{\rm lin} = 0.08$ and lower 
frequency $\omega_{\rm lin} = 13$ (blue). We highlight in gray the scales excluded in the 
analysis, i.e. $s < 20\,h^{-1}\,{\rm Mpc}$ and $s > 180\,h^{-1}\,{\rm Mpc}$. Black lines correspond 
to the monopole and quadrupole predictions of $\Lambda$CDM with power-law PPS.}
\label{fig:1D_lin}
\end{figure}

We vary the amplitude of the template 
\eqref{eqn:template} (linearly in the range ${\cal A}_{\rm lin} \in [0,\, 1]$) together with the BAO 
($\alpha_\perp$, $\alpha_\parallel$) and nuisance parameters 
($B_0$, $B_2$, $A^0_0$ $A^2_0$, $A^0_1$, $A^2_1$, $A^0_2$, $A^2_2$)$_{z_1,z_2}$
keeping fix the remaining four primordial feature parameters. 
$B_0$ and $B_2$ are used to marginalize over the power spectra amplitude, 
i.e. clustering bias amplitude and redshift-space distortions effects. All the $A^i_j$ parameters are used 
to marginalize over the broad-band effects including angle-dependent overall shape of the pwoer spectra, 
redshift space distortions, scale-dependent bias, and any errors made in our assumption of the model 
cosmology; see Ref.~\cite{BOSS:2016apd}.
These nuisance parameters cover 
also for the marginalization over effects due to the scale-dependent clustering bias contribution which is 
expected to be very small for primordial features; see \cite{Ballardini:2017qwq,Cabass:2018roz}.
We find ${\cal A}_{\rm lin} < 0.26\ (< 0.096)$ at 95\% (68\%) confidence level (CL), we show the 
marginalized poster distribution on ${\cal A}_{\rm lin}$ in Fig.~\ref{fig:1D_lin}.

The constraints are weaker compared to the one we found for the undamped linear oscillations in 
Ref.~\cite{Ballardini:2022wzu} because the best-fit frequency $\omega_{\rm lin} \simeq 14.4$ is at 
very large separation scales of the 2PCF, mostly beyond the BOSS measurements, see 
Fig.~\ref{fig:1D_lin}. Indeed, for such frequency, the best-fit amplitude ${\cal A}_{\rm lin}  = 0.16$ 
is still allowed from BOSS DR12 galaxy correlation data; note however that BOSS DR12 2PCF \cite{Ballardini:2022wzu} 
will probe more efficiently lower frequencies $\omega_{\rm lin} \lesssim 13$. 
Tighter constraints could also be obtained by full-shape matter power spectrum data \cite{Beutler:2019ojk}.

\section{Conclusions} \label{sec:conclusion}
A primordial signal with an oscillatory pattern superimposed to the PPS can mimic the smoothing excess 
in the region of the acoustic peaks measured by {\em Planck} and quantified by the phenomenological 
parameter $A_{\rm L}$.

It is important to stress that while it is possible to find a specific localized oscillation
able to reproduce a signal close to the one generated by $A_{\rm L} \simeq 1.18$ in CMB temperature 
power spectrum, we expect that this does not hold for CMB polarization \cite{Planck:2018jri}. 
Such a primordial localized oscillation can leave an imprint on the clustering of the galaxy 
distribution. 

In this paper, we have accurately modelled the imprint of this specific localized oscillation on 
the matter power spectrum.
We have run a set DM-only cosmological simulations varying some of the feature parameters of 
the template \eqref{eqn:template} at the time with $1,024^3$ DM particles in a comoving box 
with side length of $1,024\,h^{-1}\,{\rm Mpc}$ consistently with Ref.~\cite{Ballardini:2019tuc}. 
We have compared then the nonlinear CDM power spectra with the prediction from perturbation 
theory \cite{Blas:2015qsi,Blas:2016sfa} showing a good agreement when the NLO is included, 
better than 0.3\% for the envelope of the feature.
The large amplitude of the feature, i.e. ${\cal A}_{\rm lin} \simeq 0.16$, required to mimic 
the effect of $A_{\rm L} \simeq 1.18$ on the CMB temperature angular power spectrum is strongly 
damped by nonlinearities. The amplitude of the feature is reduced approximately by a factor 8 at 
$z = 0$ and it is halved at $z = 1$. 

Finally, we have derived the constraint on the amplitude keeping fixed the other feature parameters. 
We have followed the methodology developed in Ref.~\cite{Ballardini:2022wzu} to constrain 
feature templates with undamped oscillations and a localized bump with the BOSS DR12 galaxy 2PCF.
Contrary to the tight constraint found in the case of undamped linear primordial oscillations in 
Ref.~\cite{Ballardini:2022wzu}, i.e. ${\cal A}_{\rm lin}^{\rm undamped} < 0.025$ at 95\% CL, we find a 
weaker upper bound for this specific template since the position of the localized wave packet of 
linearly spaced oscillations is at very small scales, around $k\sim 0.2\ {h/\rm Mpc}$, almost outside 
the observational window of the 2PCF, see Fig.~\ref{fig:1D_lin}. Here we find a constraint of 
${\cal A}_{\rm lin} < 0.26$ at 95\% CL for $\omega_{\rm lin} \simeq 14.4$.

Future galaxy clustering measurements at high redshift $z \gtrsim 1$ from Euclid \cite{EUCLID:2011zbd} are the most sensitive to these primordial oscillations mimicking ${\cal A}_{\rm L}$ and will improve significantly the observational status presented here.

\section*{Acknowledgments}
FF would like to thank Akhil Antony, Jan Hamann, Dhiraj Kumar Hazra, Arman Shafieloo, and Matteo Tagliazucchi 
for discussions on the subject.
MB and FF acknowledge financial support from the contract ASI/ INAF for the Euclid mission 
n.2018-23-HH.0 and from the ASI Grant 2016-24-H.0 and the agreement n. 2020-9-HH.0 ASI-UniRM2 
``Partecipazione italiana alla fase A della missione LiteBIRD".

\footnotesize
\bibliographystyle{abbrv}


\end{document}